 \documentclass[final,5p,times,twocolumn]{elsarticle}

\usepackage{amssymb}
\usepackage{graphics}
\usepackage{graphicx}
\usepackage{epsfig}
\usepackage[usenames]{color}
\usepackage{longtable}
\PassOptionsToPackage{hyphens}{url}\usepackage{hyperref}
\usepackage{breakurl}
\usepackage{latexsym}
\usepackage{amsmath}
\usepackage{amsthm}
\usepackage{amsfonts}
\usepackage{dsfont}
\usepackage{bm}
\newcommand\T{\rule{0pt}{2.6ex}}
\newcommand\B{\rule[-1.2ex]{0pt}{0pt}}
\newcommand{\FC}{\;,}
\newcommand{\FD}{\;.}
\newcommand{\I}{\mathrm{i}}  %imaginary unit
\newcommand{\be}{\begin{equation}}
\newcommand{\ee}{\end{equation}}
\newcommand{\bea}{\begin{eqnarray}}
\newcommand{\eea}{\end{eqnarray}}

\newcommand{\qbar}{\overline{q}}
\newcommand{\eq}[1]{(\ref{#1})}

\journal{Physics Letters B}

\begin{document}

\begin{frontmatter}



%%%%%%%%%%%%%%%%%%%%%%%%%%%%%%%%%%%%%%%%%%%%%%%%%%%%%%%%%%%%%%%%%%%%%%%%%%%%%%%%
\title{ Predicting positive parity $B_{s}$ mesons from lattice QCD}
%%%%%%%%%%%%%%%%%%%%%%%%%%%%%%%%%%%%%%%%%%%%%%%%%%%%%%%%%%%%%%%%%%%%%%%%%%%%%%%%

\author[ugraz]{C.~B.~Lang}
\ead{christian.lang@uni-graz.at}

\author[FNAL]{Daniel Mohler}
\ead{dmohler@fnal.gov}

\author[ulub,jsi]{Sasa Prelovsek}
\ead{sasa.prelovsek@ijs.si}

\author[Tri]{R.~M.~Woloshyn}
\ead{rwww@triumf.ca}

\address[ugraz]{Institute of Physics,  University of Graz, A--8010 Graz, Austria}
\address[FNAL]{Fermi National Accelerator Laboratory, Batavia, Illinois 60510-5011, USA}
\address[ulub]{Department of Physics, University of Ljubljana, 1000 Ljubljana, Slovenia}
\address[jsi]{Jozef Stefan Institute, 1000 Ljubljana, Slovenia}
\address[Tri]{TRIUMF, 4004 Wesbrook Mall Vancouver, BC V6T 2A3, Canada}


\begin{abstract}
We determine the spectrum of $B_s$ 1P states using lattice QCD. For the
$B_{s1}(5830)$ and $B_{s2}^*(5840)$ mesons, the results are in good agreement
with the experimental values. Two further mesons are expected in the
quantum channels $J^P=0^+$ and $1^+$ near the $BK$ and $B^{*}K$ thresholds. 
A combination of quark-antiquark and $B^{(*)}$ meson-Kaon interpolating fields are used to determine the mass of two QCD bound states below the $B^{(*)}K$ threshold, with the assumption that mixing with $B_s^{(*)}\eta$ and isospin-violating decays to $B_s^{(*)}\pi$ are negligible. We predict a $J^P=0^+$ bound state $B_{s0}$ with mass $m_{B_{s0}}=5.711(13)(19)$ GeV. 
With further assumptions motivated theoretically by the heavy quark limit, a bound state with $m_{B_{s1}}= 5.750(17)(19)$ GeV is predicted in the $J^P=1^+$ channel. The results from our first principles calculation are compared to previous model-based estimates.
\end{abstract}

\end{frontmatter}


Over the years experiments have uncovered a number of mesons involving heavy
quarks that do not seem to fit the simple quark-antiquark picture suggested by
quark models. Examples of these include states in the charmonium and
bottomonium spectrum \cite{Brambilla:2014jmp} as well as the charm-strange
$D_{s0}^*(2317)$ and $D_{s1}(2460)$ \cite{Agashe:2014kda}. The latter states
are identified with the $j=\frac{1}{2}$ heavy-quark multiplet, where $j$ is the total angular momentum of the light quark \cite{Isgur:1991wq}. These were predicted to be broad states above
thresholds in potential models \cite{Godfrey:1986wj,DiPierro:2001uu,Godfrey:1985xj,Ebert:1997nk}. However, the observed $D_{s0}^*(2317)$ and
$D_{s1}(2460)$ are narrow states below the $DK$ or $D^{*}K$ thresholds
\cite{Agashe:2014kda}, and it has been suggested that the thresholds play an
important role in lowering the mass of the physical states
\cite{vanBeveren:2003kd}. In a recent lattice QCD simulation
\cite{Mohler:2013rwa,Lang:2014yfa,Lang:2014zaa} these states are seen as QCD
bound states below threshold with a mass in good agreement with experiment.

In the $B_s$ meson spectrum only two positive parity states are known from experiment
\cite{Aaltonen:2007ah,Abazov:2007af,Aaij:2012uva}, the $B_{s1}(5830)$ and
$B_{s2}^*(5840)$. The LHCb experiment should be able to see the remaining two
states ($0^+$ and $1^+$), which are expected to decay into $s$-wave states by emitting either a photon or a $\pi^0$ \cite{Bardeen:2003kt}. On the theory side there
are a number of phenomenological model and EFT mass determinations
\cite{Kolomeitsev:2003ac,Guo:2006fu,Guo:2006rp,Cleven:2010aw,Colangelo:2012xi,Cheng:2014bca,Bardeen:2003kt,DiPierro:2001uu,Ebert:2009ua,Sun:2014wea},
a determination using Unitarized EFT based on low energy constants extracted
from lattice QCD simulations \cite{Altenbuchinger:2013vwa}, and some lattice QCD
calculations in the static limit \cite{McNeile:2004rf,Foley:2006cr,Koponen:2007nr,Burch:2008qx,Michael:2010aa}. The
HPQCD collaboration has published a prediction \cite{Gregory:2010gm} taking
into account explicitly only quark-antiquark operators and
extracting only the ground states in the system. This strategy can lead to
inaccurate results in the vicinity of thresholds where meson-meson scattering can have a significant
effect. None of the previous lattice simulations clearly establish the states in question as either QCD bound states below threshold or resonances above threshold. It is this gap which we aim to fill with the current publication.

In this letter we present results for masses of the $p$-wave states of bottom-strange mesons with spin and parity quantum numbers $J^P=0^+,1^+,2^+$. For the heavy-quark doublet with $j^P=\frac{3}{2}$ masses determined using only quark-antiquark operators agree with those of the observed $B_{s1}(5830)$ and $B_{s2}^*(5840)$. This, as well as calculated mass differences between heavy-light mesons, verifies our computational setup. Then we simulate $B^{(*)}K$ scattering in the scalar (axial) channel and extract the scattering matrix. Bound state poles are found below threshold and their location determines the masses of the $B_{s0}$ and $B_{s1}$.

The gauge configurations are from the PACS-CS collaboration  \cite{Aoki:2008sm}. They have
$2+1$ flavors of dynamical quarks (up/down, strange); the bottom quark is implemented
as a valence quark. The light and strange quarks are non-perturbatively improved Wilson fermions. The lattice spacing is 0.0907(13) fm and the Pion mass
is 156(7)(2) MeV. The lattice size is $32^3\times 64$ and we use stochastic distillation \cite{Morningstar:2011ka} for the quark propagation as in our analysis of the $D_s$ mesons \cite{Mohler:2013rwa,Lang:2014yfa,Lang:2014zaa}. 
This allows to include contributions with annihilation diagrams. Further details including the $u$, $d$, and $s$ quark parameters can be found in \cite{Lang:2014yfa}.

The dynamic strange quark mass and the associated hopping parameter $\kappa_s$ used in \cite{Aoki:2008sm} differs significantly from the physical value. 
We therefore use a partially quenched strange quark $m_{s}^{val}\ne
m_{s}^{sea}$. Different determinations agree very well and yield the value
for $\kappa_s$\cite{Lang:2014yfa} which leads to the Kaon mass
$m_K=504(1)(7)$~MeV. 

The bottom quark is treated as a valence quark and the
Fermilab method \cite{ElKhadra:1996mp,Oktay:2008ex} is used. See Ref. \cite{Mohler:2012na,Lang:2014yfa} for details of our implementation. In the simplified form that we use \cite{Burch:2009az,Bernard:2010fr}, only the bottom quark hopping parameter $\kappa_b$ is tuned non-perturbatively, while the clover coefficients $c_E$
and $c_B$ are set to the tadpole improved value $c_E=c_B=c_{sw}^{(h)}=1/{u_0^3}$, where $u_0$ denotes the average  link. There are several ways of setting $u_0$  and we opt to use the Landau link on
unsmeared gauge configurations. Within this simplified approach the static
mass $M_1$ may have large discretization effects but mass differences are expected
to be close to physical \cite{Kronfeld:2000ck} and can be compared to experiment. Determining the bottom quark hopping parameter translates into determining the spin-averaged kinetic mass $M_2$ of 1S $B_s$ mesons from the lattice dispersion relation \cite{Bernard:2010fr}
\begin{align}
E(p)&=M_1+\frac{\mathbf{p}^2}{2M_2}-\frac{a^3W_4}{6}\sum_ip_i^4-\frac{(\mathbf{p}^2)^2}{8M_4^3}+ \dots\;,
\label{disp}
\end{align} 
where $\mathbf{p}=\frac{2\pi}{L}\mathbf{q}$ for a given spatial extent $L$. After trying multiple forms a simplified form without a $W_4$ term is taken\footnote{The determination of the kinetic mass $M_2$ (including its uncertainty) and thereby what is identified with the "physical" meson mass is rather insensitive (i.e. varies by $\le15\%$ of the uncertainty) to including or not including a $W_4$ term. (This is not the case for $M_4$ and its uncertainty.)} and for the value $\kappa_b=0.096$ used in our simulation we obtain $M_{2,\overline{B_s}}=5086(135)(73)$~MeV. This value is significantly smaller than the physical value $(m_{B_s}+3m_{B_s^*})/4=5403.2^{+1.8}_{-1.6}$ MeV but the effects on the binding energies used in our analysis are small.
This can be seen from the moderate  difference between  $D_s$  \cite{Lang:2014yfa} and $B_s$ binding energies we obtain and will be accounted for in the systematic uncertainty.  
For the analysis of the phase shifts the dispersion relations for the Kaon ($K$) and the heavy meson ($B$ or $B^*$) are needed. For the heavy $B$ mesons we again take Eq. \eq{disp} with $W_4=0$  and the results are tabulated in Table \ref{cdisp_rel}. For the Kaon the relativistic dispersion relation $E_K(p)=\sqrt{m_K^2+\mathbf{p}^2}$ is used.

\begin{table}[bt]
\caption{Parameter values in the dispersion relation \eq{disp} for both the
  $B$ and $B^*$ meson in lattice units. For our uncertainty estimates we also use alternate parametrizations.\label{cdisp_rel}}
\begin{center}
\small\begin{tabular}{ccc}
\hline
\T\B &  $B$&$B^*$\\
\hline
\T\B $M_1$ & 1.5742(16) & 1.5960(27)\\
\T\B $M_2$ & 2.16(29) & 2.21(43) \\
\T\B $M_4$ & 1.4(2.6) & 1.05(77) \\
\hline\end{tabular}
\end{center}
\end{table}

The discrete energy levels for our combined basis of quark-antiquark and
$B^{(*)}K$ operators are extracted from time correlations using the
variational method
\cite{Michael:1985ne,Luscher:1985dn,Luscher:1990ck,Blossier:2009kd}. For a
given quantum channel one measures the Euclidean cross-correlation matrix
$C_{ij}(t)=\langle O_i(t)O^\dagger_j(0)\rangle$  between several operators
living on the corresponding time slices. The generalized eigenvalue problem
disentangles the eigenstates $|n\rangle$. From the exponential decay of the
eigenvalues $\lambda_n(t)\sim\exp{(-E_n(t-t_0))}$ one determines the energy
values $E_n$ of the eigenstates by exponential fits to the asymptotic
behavior. The overlap factors $\langle O_i | n\rangle$ give the composition of the eigenstates
in terms of the lattice operators. In order to obtain the lowest energy eigenstates and energy levels
reliably one needs a sufficiently large set of operators with the chosen
quantum numbers. All error values come from a jack-knife
analysis, where the error analysis for the phase shift includes also the
input from the dispersion relation (\ref{disp}).

\begin{table}[tb]
\caption{Selected mass splittings (in MeV) of mesons involving bottom quarks compared to the values from the PDG \cite{Agashe:2014kda}. A bar denotes spin average. Errors are statistical and scale-setting only. \label{splittings}}
\begin{center}
\small\begin{tabular}{ccc}
\hline
Mass splitting & This work & Experiment       \cr
\hline
\T\B$m_{B^*}-m_B$ & 46.8(7.0)(0.7) & 45.78(35)\cr
\T\B$m_{B_{s^*}}-m_{B_s}$ & 47.1(1.5)(0.7)& $48.7^{+2.3}_{-2.1}$\cr
\T\B$m_{B_s}-m_{B}$ & 81.5(4.1)(1.2) & 87.35(23)\cr
\T\B$m_Y-m_{\eta_b}$ &  44.2(0.3)(0.6) & 62.3(3.2)\cr
\T\B$2m_{\overline{B}}-m_{\overline{\bar{b}b}}$ & 1190(11)(17) & 1182.7(1.0)\cr
\T\B$2m_{\overline{B_s}}-m_{\overline{\bar{b}b}}$ & 1353(2)(19)& 1361.7(3.4)\cr
\T\B$2m_{B_c}-m_{\eta_b}-m_{\eta_c}$ & 169.4(0.4)(2.4) & 167.3(4.9) \cr
\hline\end{tabular}
\end{center}
\end{table}

To test our heavy quark approach we calculate a number of mass splittings
involving heavy-light and/or heavy-heavy mesons, see Table
\ref{splittings}. The quoted uncertainties are statistical and from
scale-setting only and the values are not intended to be precision results. In
particular our lighter than physical bottom quark mass strongly affects the
spin-dependent splittings, but the effect tends to cancel with discretization errors. Estimates for both sources of uncertainty will be taken into account in our prediction of $B_s$ mesons.

Partial wave unitarity implies that the scattering amplitude $T(s)$ for
elastic $B^{(*)}K$ scattering can be written as
\be\label{eq:elscat}
\sqrt{s}\, T^{-1}(s) = p \cot \delta(s) - \I  p\FC
\ee
where $p(s)$ is the momentum and $s=E^2$ the energy squared in the center of momentum system.
Assuming a localized interaction region smaller than the spatial lattice extent a relation between the energy spectrum of meson-meson correlators in finite volume and the infinite volume phase shift $\delta$ has been derived \cite{Luscher:1985dn,Luscher:1986pf,Luscher:1990ux,Luscher:1991cf,Briceno:2014oea}, 
\begin{equation}\label{eq:luescher_z}
f(p)\equiv p \cot \delta(p)= \frac{2 \mathcal{Z}_{00}(1;(\tfrac{pL}{2\pi})^2)}{L \sqrt\pi}
\approx\frac{1}{a_0}+\frac{1}{2}r_0 p^2\FC
\end{equation}
which applies in the elastic region and in the rest frame. $\mathcal{Z}_{00}$ denotes the generalized zeta function \cite{Luscher:1990ux,Luscher:1991cf}
This real function $f(p)$ has no threshold singularity and the measured values can be found indeed above and below threshold. For $s$-wave  scattering an effective range approximation (see Eq. (\ref{eq:luescher_z})) may be used to interpolate between the closest points near threshold. The imaginary contribution to $T^{-1}$ becomes real below threshold (responsible for a cusp in $\mathrm{Re}\;T$). When the two contributions cancel, $T^{-1}$ (see Eq. (\ref{eq:elscat})) develops a zero where
\be\label{eq:pole}
f(\I|p_B|)+|p_B|=0\FD
\ee
That zero below threshold corresponds to a bound state pole of $T$ in the upper Riemann sheet.

\begin{table*}[tbh]
\caption{ Energy levels for $J^P=0^+$ (upper set), $1^+$ (middle set) and $2^+$ (lower set). A correlated 2-exponential fit is used and $\bar m=\tfrac{1}{4}(m_{B_s}+3m_{B_s^*})$ with $\bar{m}=1.62897(43)$ in lattice units. $t_0$ denotes the reference point in the generalized eigenvalue problem. Energy 2 in the middle set corresponds to the $B_{s1}^*(5830)$. The lower set shows the naive energy level for the $J^P=2^+$  and corresponds to the $B_{s2}^*(5840)$ using the same operator basis used in \cite{Lang:2014yfa} for the $D_{s2}^*$. \label{tab:A1T1_energies}}
\begin{center}
\small\begin{tabular}{c|cccccl|lccccc}
\hline
level  & $t_0$ & basis &$\textrm{fit range}$      & $\tfrac{\chi^2}{d.o.f}$ & $Ea$  & $E-\bar m~$[GeV] & $(a p)^2$  &
 $a p ~\cot(\delta)$ &$p^2$[GeV$^2$]  & $ p ~\cot(\delta)$ [GeV]  \\
\hline                                
1 & 2 & $O_{1,2,4,5,7}$  & 4-16 & 0.53  & 1.7735 (44)  & 0.315 (9)  & -0.0128 (19) &-0.106(10)& -0.0606(88) &-0.231(23)\\
2 & 2 & $O_{1,2,4,5,7}$  & 4-16 & 1.05  & 1.8213 (29)  & 0.419 (6)  &  0.0066 (13) &-0.116(18)&  0.0312(62) &-0.252(40)  \\
3 & 2 & $O_{1,2,4,5,7}$  & 3-13 & 1.35  & 1.9139 (59)  & 0.620 (13) &  0.0535 (35) &-0.045(76)&  0.2532(165)&-0.097(166) \\
\hline                                
 1 & 2 &   $O_{3,4,6,9,11}$ & 4-14  &  0.67 & 1.7919 (51) & 0.353 (11) & -0.0141(22) & -0.113(11) & -0.067(11) & -0.246(25)\\
 2 & 2 &   $O_{3,4,6,9,11}$  & 3-14  & 0.85 & 1.8255 (42) & 0.428 (9) & -- & -- & -- & --\\
 3 & 2 &   $O_{3,4,6,9,11}$  & 3-14  & 0.54 & 1.8395 (45) &0.457 (10) &  0.0050(24) & -0.142(49) &  0.024(11) & -0.308(106)\\
 4 & 2 &   $O_{3,4,6,9,11}$  & 3-14  & 1.19 & 1.9406 (50) & 0.677 (11) &  0.0566(31) &  0.021(67) &  0.268(15) &  0.046(145)\\
\hline
 1 & 2 &   $O_{1,2}$  & 4-14  & 0.43 & 1.8357(51) & 0.450 (11) & -- & -- & -- & --\\
\hline\end{tabular}
\end{center}
\end{table*}

\begin{figure}[tb]
\begin{center}
\includegraphics*[width=0.45\textwidth,clip]{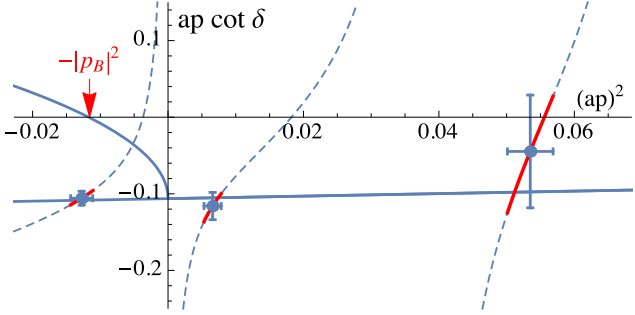}
\end{center}
\caption{Plot of $ap \cot\delta(p)$ vs. $(ap)^2$ for $BK$ scattering in $s$-wave. Circles are  values from our simulation; red lines indicate the error band following
the L\"uscher curves (broken lines). The full line gives the linear fit (\ref{eq:luescher_z}) to the points. Below threshold $|p|$ is added and the zero of the
combination (\ref{eq:pole}) indicates the bound state position in infinite volume. Displayed uncertainties are statistical only.}
\label{fig:A1_scattlen}
\end{figure}

For $J^P=0^+$  we computed cross-correlations between four $\bar{s}b$ (in the form given in Table XIII of \cite{Lang:2014yfa}) and three $BK$ (irreducible representation $A_1^+$) operators:
\begin{align}
\label{eq:Op_A1}
O_5\equiv O_1^{BK}&=\left[\bar{s}\gamma_5 u\right](\vec p=0)\left[\bar{u}\gamma_5b\right](\vec p=0)+\left\{u\rightarrow d\right\}\,,\nonumber\\
O_6\equiv O_2^{BK}&=\left[\bar{s}\gamma_t\gamma_5 u\right](\vec p=0)\left[\bar{u}\gamma_t\gamma_5 b\right](\vec p=0)+\left\{u\rightarrow d\right\}\,,\nonumber\\
O_7\equiv O_3^{BK}&=\!\!\!\!\!\!\!\!\!\sum_{\vec p=\pm e_{x,y,z}~2\pi/L}\!\!\!\!\!\!\!\left[\bar{s}\gamma_5u\right](\vec p)\left[\bar{u}\gamma_5 b\right](-\vec p)+\left\{u\rightarrow d\right\}\,,
\end{align}
where we assume that the  closeness of the $BK\pi$ threshold can be ignored for our simulation. All operators are built according to the distillation method from quark sources that
are eigenvectors of the spatial Laplacian, providing a smearing with a Gaussian-like
envelope. The gauge links are four-dimensional normalized hypercubic (nHYP) smeared \cite{Hasenfratz:2007rf}.

We omit  $B_s^{(*)}\pi$ interpolators since we work in the isospin limit where such decays cannot occur. We
also neglect $B_s^{(*)}\eta$, partially motivated by the threshold lying $\mathcal{O}$(140 MeV) above the $B^{(*)}K$ threshold. Inclusion would
necessitate a coupled channel study which would need several volumes and considerably complicate the calculation.

As in earlier experience it turned out that the full set of operators gave
noisier signals than suitable subsets so for the final analysis we use the operator set (1,2,4,5,7). The energy values resulting from correlated 2-exponential fits to the eigenvalues are given in Table \ref{tab:A1T1_energies}.

In this channel $B$ and $K$ are in $s$-wave. If there is a bound state one expects an
eigenstate with energy approaching the bound state energy from below in the
infinite volume limit. The levels above threshold then would be dominated by 
$BK$ operators with back-to-back momenta. This is exactly what is seen from the 
overlap ratios: The lowest level is dominated by operators 1,2 and 4, level 2 by the $B(0)K(0)$
operator 5 and level 3 by  the $B(1)K(-1)$ operator 7.

As shown in \eq{eq:luescher_z}  we can use the values of $p\cot\delta(p)$  
from L\"uscher's relation to determine the effective range
parametrization near threshold. The energy eigenvalues give the points shown
in Fig. \ref{fig:A1_scattlen} together with a linear fit. The value and slope at threshold
can be related to the scattering length and effective range:
\begin{align}
a_0^{BK}&=-0.85(10) \,\mathrm{fm}\;,\quad&r_0^{BK}=0.03(15) \,\mathrm{fm}\;.
\end{align}
Equation (\ref{eq:pole}) gives the bound state position. From this the binding energy is estimated to be $m_B+m_K-m_{B_{s0}}\!=\!64(13)(19)~$MeV; thus, using the physical threshold as input to minimize systematic effects, we predict a bound state $B_{s0}$ with $J^P=0^+$ at a mass of
\begin{align}
m_{B_{s0}}&=5.711(13)(19) \,\mathrm{GeV}\;.
\end{align}
The first error is due to statistics and the effective range fit, and
the second  value is our estimate for the systematic error with the main
contributions due to heavy quark discretization, unphysical Kaon mass, and finite volume effects. Details of this uncertainty estimate are provided in Table \ref{errors}.

\begin{table}[tb]
\caption{Systematic uncertainties in the mass determination of the below-threshold states with quantum numbers $J^P=0^+, 1^+$. The heavy-quark discretization effects are quantified by calculating the Fermilab-method mass mismatches and employing HQET power counting \cite{Oktay:2008ex} with $\Lambda=700$~MeV. The dominant contributions arise from mismatches in $m_B$ and $m_E$ and their size as a fraction of the reference scale $\Lambda$ can be seen in Fig. 3 of \cite{Oktay:2008ex}. The finite volume uncertainties are estimated conservatively by the difference of the lowest energy level and the pole position (see also Equations (9) and (28) of \cite{Sasaki:2006jn}).  The last line gives the effect of using only the two points near threshold for the effective range fit. 
The third point might be affected more strongly by the $B_s^{(*)}\eta$ threshold, so it is reassuring that the difference in results between two-point and three-point fits is minimal. The total uncertainty has been obtained by adding the single contributions in quadrature.\label{errors}}
\begin{center}
\small\begin{tabular}{cc}
\hline
\T\B source of uncertainty &  expected size [MeV]\\
\hline
heavy-quark discretization & 12\\
finite volume effects & 8\\
unphysical Kaon, isospin \& EM & 11\\
b-quark tuning & 3\\
dispersion relation & 2\\
spin-average (experiment) & 2\\
scale uncertainty & 1\\
3 pt vs. 2 pt linear fit & 2\\
\hline
total & 19\\
\hline\end{tabular}
\end{center}
\end{table}

For $J^P=1^+$  we computed cross-correlations between eight $\bar{s}b$ (in the form given in Table XIII of \cite{Lang:2014yfa}) and three $B^*K$ (irrep $T_1^+$) operators:
\begin{align}
\label{eq:Op_T1}
O_9\equiv O_{1,k}^{B^*K}&=\left[\bar{s}\gamma_5 u\right](\vec p=0)\left[\bar{u}\gamma_k b\right](\vec p=0)+\left\{u\rightarrow d\right\}\,,\nonumber\\
O_{10}\equiv O_{2,k}^{B^*K}&=\left[\bar{s}\gamma_t\gamma_5 u\right](\vec p=0)\left[\bar{u}\gamma_t\gamma_k b\right](\vec p=0)+\left\{u\rightarrow d\right\}\,,\nonumber\\
O_{11}\equiv O_{3,k}^{B^*K}&=\!\!\!\!\!\!\!\!\!\sum_{\vec p=\pm e_{x,y,z}~2\pi/L}\!\!\!\!\!\!\!\left[\bar{s}\gamma_5 u\right](\vec p)\left[\bar{u}\gamma_k b\right](-\vec p)+\left\{u\rightarrow d\right\}\,.\nonumber
\end{align}
Comparing various subsets of operators the most stable set was (3,4,6,9,11), where four energy
levels could be determined  (Table \ref{tab:A1T1_energies}).

Based on the overlaps, levels 3 and 4 are dominated by operators 9 ($B^*(0)K(0)$) and
11  ($B^*(1)K(-1)$), respectively. The lowest energy level (dominated by
operators 3 and 4) agrees with a bound state interpretation. 
A linear fit to the points corresponding to energy levels 1, 3 and 4 gives the scattering
parameters
\begin{align}
a_0^{B^*K}&=-0.97(16) \,\mathrm{fm}\;,\quad &r_0^{B^*K}  = 0.28(15) \,\mathrm{fm}\;.
\end{align}
This indicates a $B^*K$ bound state $B_{s1}$  with a binding energy of 71(17)(19) MeV. Using
again the physical threshold as input we obtain
\begin{align}
m_{B_{s1}}&= 5.750(17)(19) \,\mathrm{GeV}\;.
\end{align}
This state has not (yet) been observed in experiments.

Notice that our determination assumes that the effect of $s$-wave -- $d$-wave mixing is negligible on the scale of our uncertainty. This is motivated theoretically by the heavy quark limit \cite{Isgur:1991wq} (where such mixing  is absent), which should be a good approximation for bottom-strange mesons.
 
\begin{figure}[t]
\begin{center}
\includegraphics*[width=0.45\textwidth,clip]{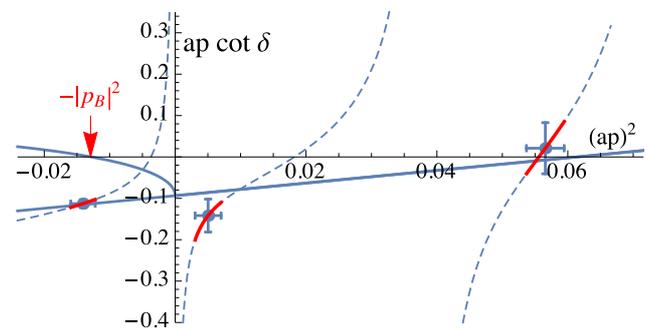}
\end{center}
\caption{Plot of $ap \cot\delta(p)$ vs. $(ap)^2$ for $B^*K$ scattering in $s$-wave, as given by the levels 1, 3 and  4 in Table \ref{tab:A1T1_energies}; see analogous caption of Fig. \ref{fig:A1_scattlen}.}
\label{fig:T1_scattlen}
\end{figure}

Level 2 (dominated by operator 6) lies just below threshold. This is interpreted, as in the case of the $D_{s1}(2536)$ \cite{Lang:2014yfa},  to be the
$j=\frac{3}{2}$ state with $J^P=1^+$ which does not couple to $B^*K$ in $s$-wave in the heavy quark limit \cite{Isgur:1991wq}. 
The composition of the state with regard to the $\qbar q$
operators is fairly independent of whether the $B^*K$ operators are included
or not. Assuming that the coupling to $B^* K$ in $s$-wave is indeed small, the
``avoided level crossing'' region is so narrow that this state may be treated as
decoupled from the $B^*K$ scattering channel. Taking the mass difference with
respect to the $B_s$ spin average and adding the physical value gives
\begin{align}
m_{B_{s1}^\prime}&= 5.831(9)(6) \,\mathrm{GeV}\FC
\end{align}
where the errors are statistical and scale-setting only. In experiments
\cite{Aaltonen:2007ah,Aaij:2012uva} one finds a resonance $B_{s1}(5830)$ decaying dominantly into $B^{*+} K^-$ 10 MeV above threshold at 5.8287(4) GeV. The masses are
in excellent agreement.

The lowest energy level
with $J^P=2^+$ (irrep $T_2^+$) 
corresponding to the $B_{s2}^*(5840)$ is extracted using just $\bar{s}b$
operators. The resulting mass is
\begin{align}
m_{B_{s2}}&= 5.853(11)(6) \,\mathrm{GeV}\FC
\end{align}
consistent with the observed value \cite{Agashe:2014kda}.

In summary we have analyzed the spectrum of positive parity $B_s$ mesons\footnote{The  binding energies of the corresponding $D_s$ mesons were also reanalyzed with our updated procedure (basis, dispersion relation, etc.) and are fully compatible with our old results \cite{Mohler:2012na,Lang:2014yfa} and, within systematic uncertainties, with experiment.} and find two bound states below threshold, corresponding to the as-yet-unobserved $B_{s0}^*$ and $B_{s1}$ 1P states. Table \ref{models} compares our first-principles lattice QCD calculation to previous results. Different variants of Unitarized ChPT along with phenomenological or lattice input (in particular \cite{Cleven:2010aw,Altenbuchinger:2013vwa}) lead to mass predictions that are in good agreement with our calculation. Also, the model based on heavy-quark and chiral symmetry by Bardeen, Eichten and Hill \cite{Bardeen:2003kt} gives results that are remarkably close.  

\begin{table}[tb]
\caption{Comparison of masses from this work to results from various model based calculations; all masses in MeV.\label{models}}
\begin{center}
\small\begin{tabular}{lcc}
\hline
$J^P$&$0^+$ & $1^+$ \cr
\hline
Covariant (U)ChPT \cite{Altenbuchinger:2013vwa} & 5726(28) & 5778(26)\cr
NLO UHMChPT \cite{Cleven:2010aw} & 5696(20)(30)& 5742(20)(30)\cr
LO UChPT \cite{Guo:2006fu,Guo:2006rp} & 5725(39)& 5778(7) \cr
LO $\chi$-SU(3) \cite{Kolomeitsev:2003ac}& 5643 & 5690\cr
HQET + ChPT \cite{Colangelo:2012xi} & 5706.6(1.2) & 5765.6(1.2) \cr
\hline
Bardeen, Eichten, Hill \cite{Bardeen:2003kt} & 5718(35) & 5765(35) \cr
\hline
rel. quark model \cite{DiPierro:2001uu}& 5804 & 5842\cr
rel. quark model \cite{Ebert:2009ua}& 5833 & 5865\cr
rel. quark model \cite{Sun:2014wea} & 5830 & 5858\cr
\hline
HPQCD \cite{Gregory:2010gm} & 5752(16)(5)(25) & 5806(15)(5)(25)\cr
this work &  5713(11)(19) & 5750(17)(19)\cr
\hline\end{tabular}
\end{center}
\end{table}

\section*{Acknowledgements}

We thank the PACS-CS collaboration for providing the gauge configurations. D.~M. would like to thank E.~Eichten, R.~Van~de~Water and J.~Simone for insightful discussions. The calculations were performed on computing clusters at the University of Graz (NAWI Graz) and with USQCD resources at Fermilab, supported by the DOE. This work is supported in part by the Austrian Science Fund (FWF):[I1313-N27], by the Slovenian Research Agency ARRS project N1-0020 and by the Natural Sciences and Engineering Research Council of Canada. Fermilab is operated by Fermi Research Alliance, LLC under Contract No. De-AC02-07CH11359 with the United States Department of Energy. 

\section*{\small References}

\bibliographystyle{elsarticle-num}
\bibliography{Lgt}

\end{document}